\documentclass[twocolumn,preprintnumbers,amsmath,amssymb,prb]{revtex4}
\usepackage[dvips]{graphicx}
\usepackage{dcolumn}
\usepackage{bm}

\begin{document}


\title{EXTRAORDINARY PHENOMENA IN SEMICONDUCTOR-METAL HYBRID NANOSTRUCTURES
 BASED ON BILINEAR CONFORMAL MAPPING
}

\author{S. A. Solin}
\affiliation{Center for Materials Innovation
And Department of Physics, Washington University in St. Louis 
1 Brookings Drive St. Louis, Missouri, 63130, USA}%

\maketitle
\section{INTRODUCTION}
The recent discovery by Solin and co-workers of the phenomenon of 
extraordinary magnetoresistance (EMR) in hybrid structures of narrow-gap 
semiconductors and non-magnetic metals \cite{Solin:2000} portends 
considerable promise for impacting magnetic sensor technology for 
macroscopic, microscopic and nanoscopic applications. \cite{Solin:2001} 
It has been recently realized by Solin and his colleagues that the EMR 
phenomenon, is but one example of a broad class of geometry-based 
interfacial effects in hybrid structures formed from the juxtaposition of a 
semiconductor such as InSb, InAs \cite{Moller:2002} or 
Hg$_{1-x}$Cd$_{x}$Te \cite{Solin:1996} with a metal such as Au. Indeed, 
one can envision a number of other "EXX" interfacial effects in metal 
semiconductor hybrid structures where E = extraordinary and XX = 
piezoconductivity, optoconductivity, electroconductivity, thermoconductivity 
etc. and combinations of these such as magnetothermoconductivity. Both EPC 
\cite{Rowe:2003} and EOC \cite{Wieland:1} have recently been 
demonstrated.

The EXX effects form the basis for a number of new macroscopic and 
nanoscopic sensor devices analogous to those such as automobile ignition 
sensors and ultra-high-density read-head sensors, respectively, that derive 
from the discovery of the EMR phenomena. Several of the applications of EXX 
sensors that one can envision, especially in the field of medicine, require 
scaling of an EXX structure to nanoscopic dimensions. This has already been 
achieved in the case of EMR sensors and devices with dimensions of order 20 
nm have been shown to exhibit new mesoscopic physics effects and to exhibit 
very high values of EMR. \cite{Solin:2002} However, the underlying 
principal on which the scaling of internally shunted macroscopic EXX 
structures to externally shunted working devices in the nanoscopic size 
regime has to date not been fully elucidated. Accordingly, we describe here 
the bi-linear conformal mapping procedure that has been applied to the 
scaling of EMR devices and show how that procedure can be applied to EXX 
structures in general.

\section{BACKGROUND}

There are two principal contributions to the magnetoresistance of any 
resistive device, namely a physical contribution and a geometric 
contribution. \cite{Popovic:1991} The physical contribution derives from 
the dependence of intrinsic material properties such as carrier 
concentration and carrier mobility on the applied magnetic field. The 
geometric contribution is an extrinsic property that depends on the shape of 
the device, the placement and geometry of the (metallic) contacts and, the 
placement and geometry of any inhomogeneities or shunts that may be present. 
The geometric contribution to the MR also depends on the intrinsic physical 
properties of the inhomogeneities relative to those of the host material, 
e.g. on the conductivity ratio. \cite{Tineke:1998} For most materials of 
current interest as MR sensors such as layered magnetic metals which exhibit 
giant MR (GMR) \cite{Egelhoff:1995} or tunnelling MR (TMR) 
\cite{Mitra:2001} and the magnetic layered oxide manganites which exhibit 
colossal MR (CMR), \cite{Jin:1994} the physical contribution to the MR is 
dominant. However, Solin and his colleagues have shown that judiciously 
designed hybrid structures composed of a non-magnetic narrow-gap 
semiconductor matrix with high carrier mobility and a non-magnetic metallic 
inhomogeneity or shunt can exhibit a room temperature MR that is not only 
dominated by the geometric contribution but also attains room temperature 
values of order 1,000,000{\%}. \cite{Solin:2003} This is several orders 
of magnitude higher than what has been achieved with conventional GMR, TMR 
or CMR devices. The new phenomenon was subsequently dubbed extraordinary MR 
or EMR. \cite{Zhou:2000}

The proof of principal demonstration of EMR was accomplished with symmetric 
4-probe macroscopic van der Pauw (vdP) disc structures formed from Te- doped 
InSb (electron concentration $n=2\times10^{17}$ cm$^{-3}$
and mobility $\mu=4.5\times10^{4}$ cm$^{2}$/Vs) containing a concentric cylindrical metallic inhomogeneity as depicted in 
the inset of Fig. 1. Solin \textit{et al.} also showed \cite{Solin:2004} that in 
general,
\begin{eqnarray}
EMR(\Delta H, H_{bias})=\nonumber\\
 \frac{R^{eff}(\Delta H + H_{bias})-R^{eff}(H_{bias})}{R^{eff}(H_{bias})}
\end{eqnarray}
where $H$ is the applied field normal to the plane of the device, 
$R^{eff}(H)$ is the effective field-dependent resistance measured in a 4-probe 
configuration $H_{bias}$ is the bias field and $\Delta H$ is the applied or signal 
field (not the field gradient). In the zero bias large signal but low field 
limit, $\mu\Delta H\ll 1$,
\begin{eqnarray}
EMR(\Delta H, 0)= \frac{R^{eff}(\Delta H)-R^{eff}_{0}}{R^{eff}_{0}}=\nonumber\\
G_{S}(\Delta H)[\mu \Delta H]^{2}\pm G_{AS}(\Delta H)[\mu \Delta H].
\end{eqnarray}
Here $G_{S}(\Delta H)$ and $G_{AS}(\Delta H)$ are, respectively, symmetric and antisymmetric 
geometric factors that depend on the shape, location and physical properties 
of the conducting inhomogeneity and contacts while 
$R_{eff}(0)=R_{0}^{eff}$. [For the symmetric structure shown in the inset of 
Fig. 1. $G_{AS}(\Delta H)=0$.] Clearly, narrow-gap high mobility semiconductors such as 
InSb are choice materials for EMR devices.
\begin{figure}
\includegraphics[width=0.5\textwidth]{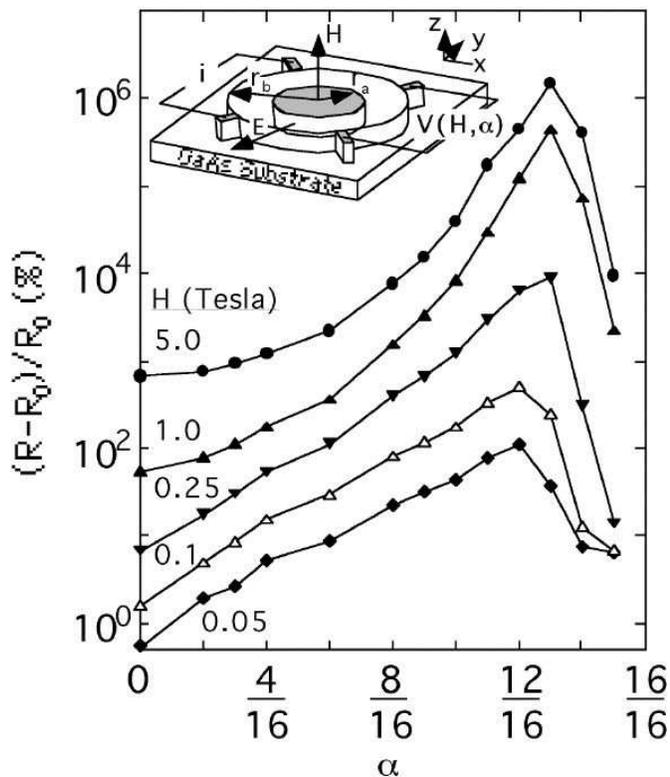}
\caption{The field-dependence of the magnetoresistance, 
$(R-R_{0})/R_{0}$, of a Te-doped InSb van der Pauw disk of radius $r_{b}$ in which 
is embedded a concentric right circular cylinder of Au of radius $r_{a}$. The 
filling factor is $\alpha =r_{a}/r_{b}$. Inset:-- a schematic diagram of the hybrid 
disk structure.}
\end{figure}

The magnetotransport properties of the macroscopic vdP structure shown in 
the inset of Fig. 1. can be quantitatively accounted for using the above 
equations together with both finite element analysis \cite{Moussa:2001} 
and analytic techniques. \cite{Zhou:2001} However, the EMR phenomenon can 
also be readily understood using a simple though non-intuitive classical 
physics analysis. The components of the magnetoconductivity tensor 
\underline{$\sigma (H)$} for the semiconductor are $\sigma _{xx}(\beta )=\sigma _{yy}(\beta )=\sigma /[1+\beta ^{2}]$,
$\sigma _{zz}(\beta )=\sigma $ and 
$\sigma _{xy}(\beta )=-\sigma _{yx}(\beta)=-\sigma \beta /[1+\beta ^{2}]$ with $\beta =\mu H$ and all
other tensor components being 
zero. If the electric field on the vertical surface of the inhomogeneity is 
$\textbf{E}=E_{x}\textbf{x}+E_{y}\textbf{y}$, the 
current density is $\textbf{J}=\underline {\sigma 
(H)}\textbf{E}$. The electric field is everywhere normal to the 
equipotential surface of a highly conducting inhomogeneity. At $H = 0$, \underline 
{$\sigma (H)$} is diagonal so $\textbf{J}=\sigma \textbf{E}$ 
and the current flows into the inhomogeneity which acts as a \textit{short circuit}. At high $H$ 
($\beta > 1$), the off-diagonal components of \underline {$\sigma (H)$} 
dominate so $\textbf{J}=(\sigma /\beta )[E_{y}\textbf{x}-E_{x}\textbf{y}]$, 
and\textbf{ J}$\bot $\textbf{E}. Equivalently, the Hall 
angle between the electric field and the current density approaches 
90$^{o}$, and the current becomes tangent to, i.e. deflected around, the 
inhomogeneity. Thus, the inhomogeneity acts as an \textit{open circuit}. The transition of the 
inhomogeneity from short circuit at low $H$ to open circuit at high $H$ results in 
a geometric enhancement of the MR of the semiconductor even if its 
resistivity (conductivity) is field-independent (i.e. the physical MR is 
zero).

Unfortunately, the internally shunted EXX structure shown in the inset of 
Fig. 1 is not conducive to fabrication on the nanoscopic scale because it is 
difficult if not impossible to properly embed or fill a nanoscopic hole with 
metal while maintaining good electrical contact with the quasi-vertical 
sidewall. Fortunately, as we now demonstrate, one can construct an 
externally shunted metal-semiconductor hybrid structure that is not only 
galvanomagnetically equivalent to the circular structure shown in the inset 
of Fig. 1, but is also ``fabrication friendly'' on the nanoscale.

\section{CONFORMAL MAPPING}
It is known \cite{Popovic:1992} that any homogeneous device with a 
circular boundary of unit radius in the imaginary two dimensional complex 
$t$-plane with orthogonal axes $r$ and $is$ where $t = r + is$ can be mapped into the 
complex upper half Cartesian $z$-plane with orthogonal axes $x$ and $iy$ where $z = 
x + iy$ [see Fig. 2a)] by using the bilinear transformation 
\cite{Popovic:1992}
\begin{eqnarray}
z(t)=-i\frac{t+i}{t-i}
\end{eqnarray}
The above mapping equation transforms the four symmetrically spaced 
electrical contacts on the perimeter of the disk in the $t$-plane [shown in 
Fig. 2a) in the configuration for a magnetoresistance measurement] to the 
corresponding contacts on the line $y = 0$ in the $z$-plane. Although the mapped 
contacts are symmetric about the line $x = 0$ they are not of equal size as 
they are when viewed in the $t$-plane. If one embeds an off-centered hole of 
radius r$_{1}$ into the homogeneous disk of

\begin{figure}
\includegraphics[width=0.5\textwidth]{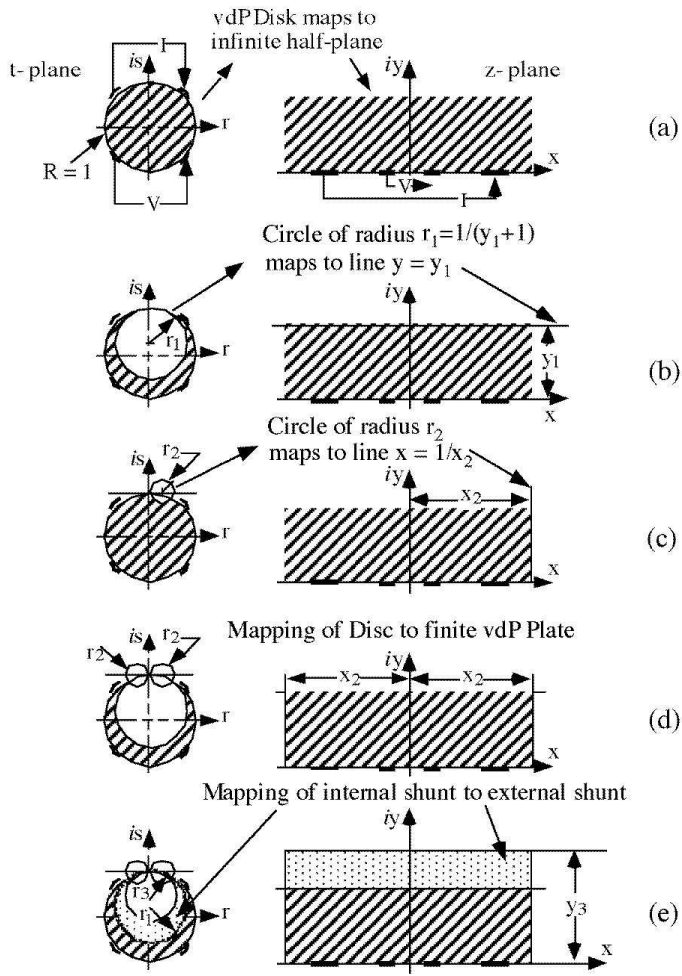}
\caption{Schematic diagram of bilinear mapping of a) a van der 
Pauw disk in the $t$-plane mapped to the upper half space of the $z$-plane. Also 
shown is the contact configuration for measuring magnetoresistance. b) an 
off-center hole in the $t$-plane mapped to a line in the $z$-plane, c) a 
circular perimeter cut in the $t$-plane mapped to a line in the $z$-plane, d) 
repeat of c) with a symmetric perimeter cut, e) an internal shunt in the 
$t$-plane mapped to an external shunt in the $z$-plane. Note: panels a) - d) 
follow Reference \onlinecite{Popovic:1991}, p.163.
Fig. 2a) as shown in Fig. 2b), that hole maps to a line that truncates the 
upper half plane at height $y_{1} = 1/(r_{1} + 1)$. In other words, the 
vacuum inside the hole of radius r$_{1}$ in the disk in the $t$-plane maps to 
the vacuum above the line $y_{1}$ in the $z$-plane. Consider now the circle of 
radius $r_{2}$ which creates an evacuated notch in the disk in the $t$-plane 
as shown in Fig. 2c). That circle maps to a line which truncates the $z$-plane 
at the position $x = x_{2} = 1/r_{2}$ as is also shown in Fig. 2c). A 
symmetrically displaced circle of equal radius on the left of the vertical 
bisector of the disk in the $t$-plane truncates the $z$-plane with a line at 
position $x_{2} = -(1/r_{2})$ as shown in Fig. 1d). By a selection of 
circular cuts in the $t$-plane, the truncated disk can be exactly mapped to a 
rectangular structure of appropriate dimension in the $z$-plane.}
\end{figure}

Of the structures depicted in Figs. 2a) - 2d), that shown in Fig 2b) is the 
simplest one which contains a fully enclosed inhomogeneity, e.g. a circular 
hole displaced from the center of the disk. If we embed this hole with a 
highly conducting metal, then the resultant structure which we call an 
off-center vdP disk is similar to the centered vdP disk which yielded the 
large EMR values cited above. However, the corresponding rectangular mapped 
structure in the $z$-plane would be of infinite extent in the $+x$ and $-x$ 
directions and would contain an external shunt of infinite height in the $+y$ 
direction. To avoid these complications, we define a structure which 
contains not only the $r_{2}$ cuts of Fig. 2d) but also an additional circle 
of radius $r_{3}$ in the $t$-plane as shown in Fig. 2e). The latter maps to 
the line $y = y_{3}$ in the $z$-plane. The modified off-centered vdP disk now 
contains a metallic inhomogeneity embedded into the space between the 
circles of radii $r_{1}$, $r_{2}$ and $r_{3}$ while the space between the 
circle of radius $r_{1}$ and the disk perimeter contains a narrow-gap 
semiconductor. Thus, the $t$-plane disk with an INTERNAL embedded shunt maps 
to a rectangle in the $z$-plane with a corresponding EXTERNAL metallic shunt. 
Moreover, for the exact mapping depicted in Fig. 2 e), the electrical 
behavior of the two structures will be identical. \cite{Popovic:1991}

Although the mapping technique described above has been known \cite{Popovic:1991}, the 
adaptation of this technique to the design of rectangular structures with 
external shunts has not been previously considered. Furthermore, for mapped 
plates with $x_{2}> 4$, the cuts represented by the circles of radius 
$r_{2}$ in the left panel of Fig. 2e) are small/negligible. Therefore, the 
externally shunted plate structure shown on the right panel of Fig 2e) is, 
to a good approximation, electrically equivalent to the vdP disk shown in 
the left panel of Fig. 2e) without the $r_{2 }$ cuts. Moreover, an expression 
for the filling factor of the rectangular EXX structure, though more complex 
than that of the concentric circular structure has been derived as a 
function of the geometric properties of the structure. 
\cite{Zhou:2001}

\section{EXPERIMENTAL CONFIRMATION OF THE MAPPING PROCEDURE}

In order to confirm that the mapping procedure described above is viable we 
test it using macroscopic EMR structures. The EMR of a non-biased 
macroscopic (long dimensions $\sim 5 $ mm) vdP plate formed from InSb and Au 
is shown in Fig. 3 as a function of the geometry of the metallic shunt and 
of the placement of the current and voltage leads. The material constituents 
of the externally shunted rectangular
\begin{figure}
\includegraphics[width=0.5\textwidth]{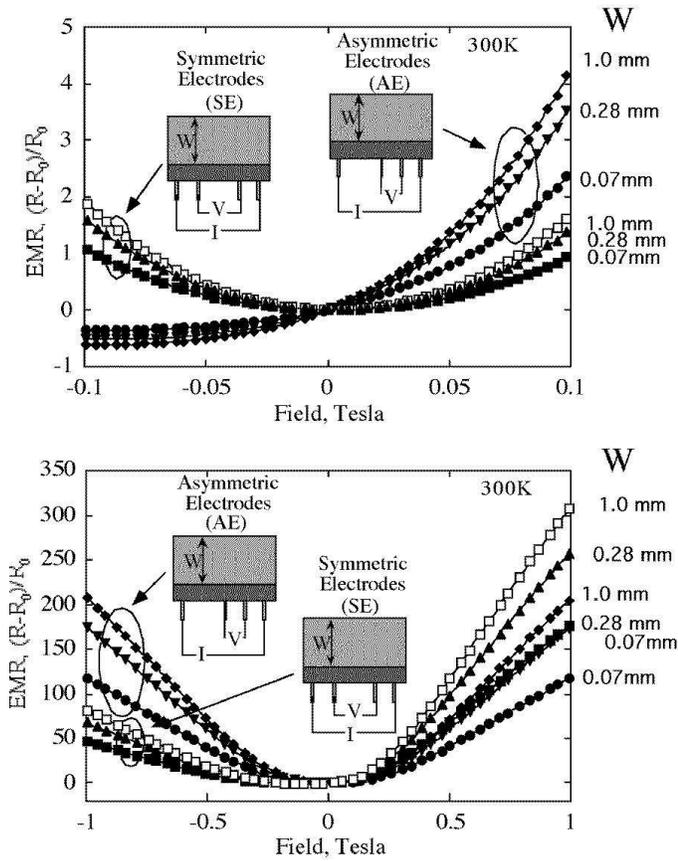}
\caption{The low field (upper panel) and high field (lower panel) room temperature EMR of van der Pauw plates with
symmetric and asymmetric electrode configurations as a function of the width, $W$, of the external shunt.  The dark
(light) rectangle represents Te:InSb (Au).}
\end{figure}
semiconductor-metal hybrid structures depicted in Fig. 3 were the same as 
those used for the internally shunted circular hybrid structures depicted in 
Fig. 1. Two features are noteworthy in the data of Fig. 3.: a) the 
room-temperature EMR is very large reaching values as in excess of 100{\%} 
at a field of 500 Gauss. b) The EMR is asymmetric with respect to the 
applied field when the leads are placed asymmetrically on the rectangular 
narrow-gap-semiconductor plate. The latter feature constitutes a condition 
of self-biasing which is important for a number of applications in which the 
sign of the applied magnetic field must be determined.

The magnetotransport properties of the externally shunted vdP plate clearly 
depend strongly on the placement of the current and voltage leads, e.g. on 
lead geometry. These properties also depend strongly on the shape and 
relative dimensions of the semiconductor and metal components of the hybrid 
structure itself as can be seen from the dependence of the EMR in Fig. 3 on 
the width, $W$, of the shunt for a semiconductor region of fixed width. Note 
that in the range of filling factors addressed in Fig. 3 which is below the 
optimum filling factor of 13/16 as depicted in Fig. 1, the EMR 
systematically increases with increasing filling factor. In addition, an 
added feature of the data of Fig. 3 is the enhancement of the EMR in the 
structure with asymmetric leads relative to the structure with symmetric 
lead placement.

\section{VALIDATION OF MAPPING TO THE NANOSTRUCTURE REGIME}
Having established that extremely high values of EMR can be obtained from 
macroscopic semiconductor-metal hybrid structures, we now address the 
challenge of scaling such EMR devices to the nanoscopic sizes required for 
ultra-high-spatial resolution and high sensitivity detection of magnetic fields. To achieve high 
spatial resolution in the vertical direction (normal to the plane of the EMR 
structure) it is necessary to use ultra thin semiconductor films. 
\begin{figure}
\includegraphics[width=0.5\textwidth]{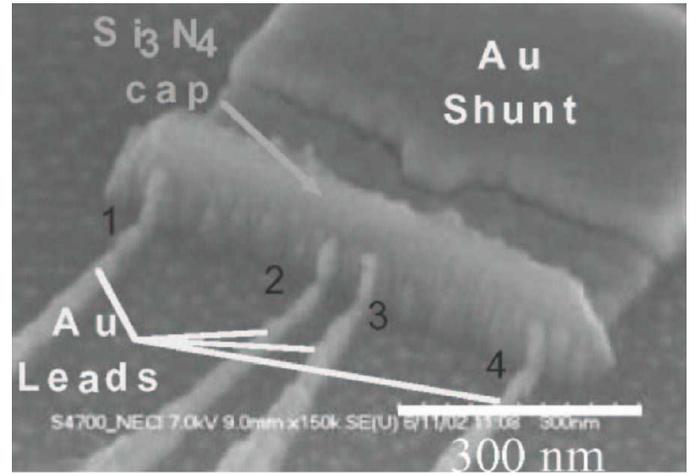}
\caption{An electron micrograph of a hybrid nanoscopic van der Pauw EMR plate structure fabricated from an
InSb/In$_{1-x}$Al$_{x}$Sb quantum well heterostructure.  The current leads, voltage leads and external shunt are labeled as
indicated.  The four contacts shown in the micrograph extend along the mesa floor and up the side of the mesa to the
upper 25 nm Al$_{0.15}$In$_{0.85}$Sb barrier.}
\end{figure}
Homogeneous InSb on GaAs is not suitable for this purpose because 
dislocations at the semiconductor substrate interface cause a drastic 
reduction in the carrier mobility, and concomitantly the EMR which scales 
with the square of the mobility, when the thickness of the semiconductor 
drops below 1 $\mu$m. \cite{Stradling:1991} To overcome this, Solin and 
coworkers used a quantum well structure, InSb/InAl$_{1-x}$Sb$_{x}$, and 
state of the art suspended mask e-beam lithography incorporating a new type 
of resist, calixarine, to fabricate the structure shown in Fig. 4. 
\cite{Solin:2003} Details of the fabrication method are provided 
elsewhere. \cite{Solin:2002} Note that the structure shown in Fig. 4 
is a nanoscopic version of the structure depicted in Fig. 2(e).
\begin{figure}
\includegraphics[width=0.5\textwidth]{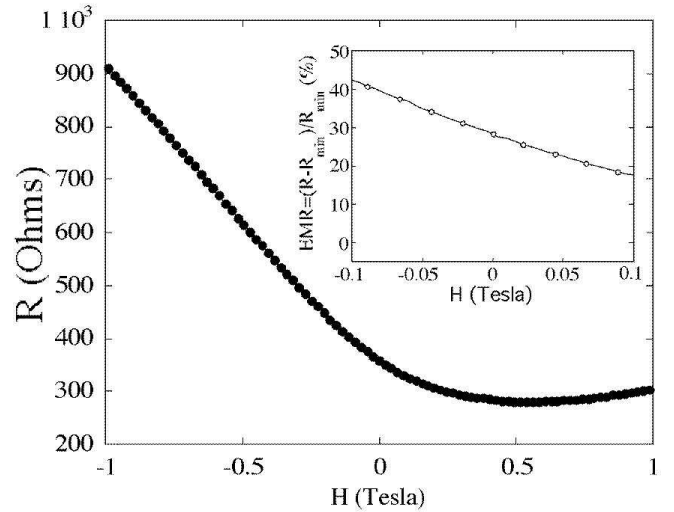}
\caption{The field dependence of the of the resistance of the 
hybrid nanoscopic van der Pauw plate structure shown in Fig. 4. Inset - the 
extraordinary magnetoresistance (referenced to $R_{min})$ of the hybrid 
nanoscopic van der Pauw plate in the low field region.}
\end{figure}

The field dependence of the room temperature magnetoresistance of the 
externally shunted nanoscopic EMR device shown in Fig. 4 is shown in Fig. 5. 
As can be seen, the EMR reaches values as high as 5 {\%} at a signal field 
of 0.05 T. To our knowledge, this is the highest room temperature MR level 
obtained to date for a patterned magnetic sensor with this spatial 
resolution. Moreover, with a modest bias field of 0.2 T corresponding to the 
zero-field offset \cite{Solin:1997} in Fig. 5, the measured EMR is 35{\%} 
at a signal field of 0.05 T. [The offset is associated with the asymmetric 
placement of the leads.] Also note that the device can be biased into a 
field region where the EMR response is linear with field, a feature that can 
simplify signal amplification. Equally significant is the fact that the 
current sensitivity, at a magnetic field bias of 0.2 T has a large measured 
value of 585 $\Omega$/T at room temperature. It is this figure that enters 
directly into the calculation of the power signal to noise ratio 
\cite{Weissman:1988} which is found from the data of Fig. 5 to be 44.5 dB 
for a bandwidth of 200 MHz and a signal field of 0.05 T at a bias of 0.2 T. 
[Note: It is common but incorrect practice in some parts of the layered 
metals community to use MR as the figure of merit but the proper measure 
should be the power signal to noise ratio at the operating bandwidth. The 
latter contains the MR but depends on other factors as well.]

\section{EXTENSION OF MAPPING TO OTHER EXX PHEMOMENA}

As noted in the introduction, EMR is but the first example of a general 
class of phenomena collectively referred to as EXX phenomena. To elucidate 
this point consider Fig. 6 which shows a three dimensional view of the 
hybrid structure depicted in Fig. 2(e). 
\begin{figure}
\includegraphics[width=0.5\textwidth]{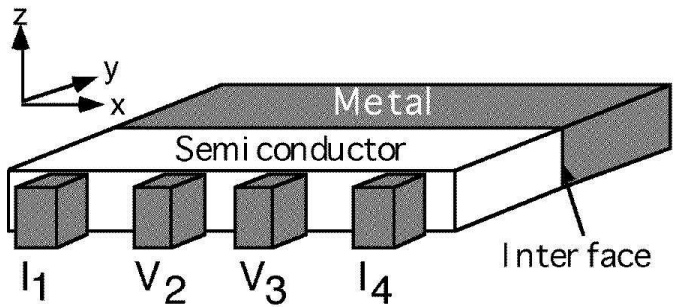}
\caption{A schematic diagram of a hybrid semiconductor-metal structure configured for a 4-probe measurement.  $I$ and $V$
refer, respectively to current and voltage probes.}
\end{figure}
The 4-probe effective resistance of this composite structure is $R_{eff} =V_{23}/I_{14}$, 
where $I$ and $V$ represent current and voltage probes, respectively. Now 
$R_{eff}$ will depend on the relative conductivities of the metal and 
semiconductor (typically $\sigma _{metal}/\sigma _{semiconductor}  >  1000$), on the 
interface resistance between them and on the specific placement of the 
current and voltage probes (e.g. on the geometry). In a non-perturbed state, 
the highly conducting metal acts as an effective current shunt, provided 
that the interface resistance is low, and $R_{eff}$ of the composite 
structure can be close to that of the metal. If a relatively small external 
perturbation such as an applied magnetic field, electric field, strain, 
temperature change etc. significantly alters the interface resistance and, 
thus, the current flow across the interface, this alteration can manifest 
itself as a large change in the effective resistance or, equivalently, in 
the output voltage (signal) of the ``sensor'' for that particular external 
perturbation.

We have illustrated this point with the EMR example where we understand the 
detailed physics of the mechanism by which the applied field perturbs the 
interface resistance. However, from an empirical viewpoint, other 
perturbations such as a strain field, photon field, etc. which perturb the 
interface resistance will yield a signal (voltage response) that is 
geometrically amplified by the hybrid structure. Thus the fabrication 
advantage for preparing an EMR nanostructure with desirable properties, is 
also applicable to other EXX phenomena when mapped to nanostructures of the 
type shown in Fig. 6.

\section{CONCLUSIONS}

We have shown that bilinear conformal mapping can be used to transform 
4-lead internally shunted EMR semiconductor-metal hybrid structures to 
galvanomagnetically equivalent externally shunted 4 lead structures. The 
latter are compatible with the fabrication of nanoscale EMR devices while 
the former are not. Mapped rectangular EMR van der Pauw plate exhibit very 
large EMR values in both macroscopic and nanoscopic form. We have also shown 
that the mapping procedure applied in the case of EMR will also be 
applicable to other generalized EXX structures.
\section{ACKNOWLEDGMENTS}
This work was supported by the U.S. National science foundation under grant 
ECS-0329347. I grateful acknowledge useful collaboration with A.C.H. Rowe, 
D.R. Hines and T. Zhou.

\end{document}